# Transferable Coarse-Grained Model for Methacrylate-Based Copolymers

Gerardo Campos-Villalobos, Flor R. Siperstein, and Alessandro Patti*

*School of Chemical Engineering and Analytical Science, The University of Manchester, Sackville Street, M13 9PL, Manchester, UK*

E-mail: Alessandro.Patti@manchester.ac.uk

## Abstract

A versatile and transferable coarse-grained (CG) model was developed to investigate the self-assembly of two ubiquitous methacrylate-based copolymers: poly(ethylene oxide-*b*-methylmethacrylate) (PEO-*b*-PMMA) and poly(ethylene oxide-*b*-butylmethacrylate) (PEO-*b*-PBMA). We derive effective CG potentials that can reproduce their behaviour in aqueous and organic polymer solutions, pure copolymer systems, and at the air-water interface following a hybrid structural-thermodynamic approach, which incorporates macroscopic and atomistic-level information. The parameterization of the intramolecular CG potentials results from matching the average probability distributions of bonded degrees of freedom for chains in solution and in pure polymer systems with those obtained from atomistic simulations. Potential energy functions for the description of effective intra- and intermolecular interactions are selected to be fully compatible with the MARTINI force-field. The optimized models allow for an accurate prediction of the structural properties of a number of methacrylate-based copolymers of different length and at different thermodynamic state points. In addition, we propose a single-segment model for tetrahydrofuran (THF), an organic solvent commonly used in methacrylate-based polymer processing. This model exhibits a fluid phase behaviour qualitatively close to that predicted by more sophisticated molecular models and can reproduce the experimental free energy of transfer between water and octanol.

## 1. Introduction

Recognizing the rich variety of opportunities deriving from the bottom-up self-assembly of amphiphilic molecules in solution has allowed the synthesis of porous materials with uniform and tunable pore dimensions and large surface areas.[1,2] These materials, resulting from the cooperative templating with a silica precursor, have been successfully applied in catalysis,[3] adsorption[4] and clean energy technologies, such as biofuels production.[5] Crucial for their synthesis is the fundamental understanding of the phase and aggregation behaviour of the amphiphilic building blocks, which determines the existence of suitable mesophases and controls their internal organization.

Amphiphilic diblock copolymers are macromolecules consisting of two connected blocks, which exhibit very different behaviour in a solvent: one block is solvophilic and can be solvated by the solvent, while the other is solvophobic. If the solvent is water, the former is referred to as the hydrophilic block and the latter as the hydrophobic block. This functional ambivalence with the solvent explains why block copolymers have acquired a central role in nanomaterials science and engineering. Specifically, in order to limit the contact between their solvophobic blocks and the solvent, at sufficiently large concentrations a number of copolymer's molecules aggregate into complex nanostructures, where solvophilic and solvophobic moieties are clearly separated. This spontaneous aggregation or self-assembly, which has captured the interest of a wide scientific community for its tremendous impact in *e.g.* nanomedicine (drug delivery and nanoreactors),[6] materials science (templated synthesis of porous solids),[7] and environmental protection (water remediation),[8] stems from the ceaseless competition between enthalpic and entropic contributions to the system free energy. This delicate equilibrium, which determines the morphology of the aggregates, can be easily al-tered by almost intangible thermal fluctuations of few $k_B T$, with $T$ the absolute temperature and $k_B \simeq 1.38 \times 10^{-23}$ J K$^{-1}$ the Boltzmann constant.

At the critical micellar concentration (CMC), isolated amphiphilic molecules start to merge together in aggregates of spherical or cylindrical shape, usually referred to as micelles, consisting of a solvophobic core surrounded by a solvophilic corona. Block copolymers have also been observed to form vesicles (or polymersomes),[9,10] being hollow, lamellar structures resulting from the interdigitation of the solvophobic moieties of two curved monolayers. At larger copolymer concentrations, well above the CMC, mesophases with a significant degree of internal organization, such as cubic, hexagonally-packed, and lamellar liquid crystals, can form. Bicontinuous phases, where the solvophilic and solvophobic moieties are clearly separated, but do not necessarily show a significant long-ranged order, have also been observed in experiments,[11] predicted by theory[12] and by molecular simulations.[13]

The specific shape of the aggregates dramatically depends on the architecture of their molecular building blocks, precisely on the length ($l$) and volume ($v$) of the solvophobic block, and on the specific aggregate/solvent interface area ($a$). The latter contribution directly depends on excluded-volume effects between neighbouring solvophilic blocks and their interactions with the solvent. This dependence is concisely expressed by the molecular packing parameter $p = v/al$, proposed by Israelachvili and co-workers in the 1970s.[14] In particular, spherical and cylindrical micelles should form for $0 \leq p \leq 1/3$ and $1/3 \leq p \leq 1/2$, respectively, while vesicles are expected in the range $1/2 \leq p \leq 1$. Basically, if one is able to estimate the packing parameter, then aggregate shape and size can be easily predicted. However, for many amphiphilic molecules the ratio $v/l$ does not depend on the tail length and can be assumed to be a constant,[15] thus limiting to the aggregate/solvent interface area the dependence of the packing parameter and disregarding the effect of the solvophobic block. Successive studies have clarified that this is not the case as both entropic and enthalpic contributions of the solvophobic group are key to determine the equilibrium aggregate structure.[16]

Block copolymers incorporating methacrylate-based blocks, such as poly(ethylene oxide-*b*-methylmethacrylate) (PEO-*b*-PMMA) and poly(ethylene oxide-*b*-butylmethacrylate) (PEO-*b*-PBMA), are currently being employed in a wide spectrum of practical applications, including the preparation of macroporous polymer electrolytes[17,18] and nanoporous membranes,[19,20] and in the templated synthesis of ordered porous solids.[21–23] In particular, by employing blends of poly(vinylidene fluoride) (PVDF) and PEO-*b*-PMMA, Zhang and co-workers obtained macroporous polymer electrolytes that exhibit very good electrochemical stability and high ionic conductivity at ambient temperature, properties that make these materials especially suitable for applications in lithium ion batteries.[17] PEO-*b*-PBMA has been used to fabricate PVDF ultrafiltration membranes with enhanced separation performance and resistance to fouling.[20] The authors indicate the presence of copolymer micelles, induced by a PEO-metal ion complex, as the main factor causing an increased porosity in the selective membrane layer. The presence of these micelles, however, would not fully explain the better anti-fouling properties, which are most probably due to the organization of the copolymer's hydrophilic blocks at the membrane/water interface.

Methacrylate-based block copolymers have also been observed to self-assemble in water-miscible organic solvents (*e.g.* THF, DMF and DMSO) into micelles, multi-lamellar vesicles and bicontinuous polymer nanospheres (BPNs), upon slow addition of water.[24] BPNs have been recognized as excellent supramolecular building blocks for the templated synthesis of porous materials incorporating specific distributions of micro-, meso-, and macropores.[21–23] Sommerdijk and co-workers investigated the effect of the relative length of PEO and PBMA blocks and their degree of affinity with the organic solvent on the formation of BPNs.[25] Their study revealed an intriguing change of morphology, from BPNs to lamellae, when dioxane, rather than THF, was employed as nonselective organic solvent.

One can therefore infer that a full insight into the solvent/copolymer interactions is crucial to predict and control the formation of the desired mesophase. Nevertheless, unravelling the nature of these interactions and their relative impact on the self-assembly process is anything but trivial, especially because of their multifaceted nature, which harmoniously determines the final morphology of the aggregates. By precisely controlling the physico-chemical parameters and boundary conditions of a self-assembly process, molecular simulation can be especially effective to overcome this challenge and provide a clear insight into the physical laws underpinning the associated kinetics and equilibrium.

Atomistic models are generally too computationally demanding to determine the behaviour of matter beyond the nanometre and nanosecond scales. Therefore, a simplified or coarse-grained (CG) approach, where a number of interactions sites are merged together, is key to enhance our ability to disclose the self-assembly pathway by molecular simulation. Lattice CG models have been successfully applied to investigate the phase behaviour of diblock copolymers in dilute[13,26,27] and concentrated[28,29] aqueous solutions. Larson and co-workers proposed the use of a lattice model to investigate the self-assembly of block copolymers in an implicit solvent and observed the formation of different liquid crystals, including hexagonal and lamellar phases.[13] This model was then modified to study ternary systems containing also an inorganic precursor to mimic the templated synthesis of hybrid organic-inorganic materials.[30–34] Although these models allow for a qualitative understanding of the self-assembly process and can reproduce the formation of micelles and liquid crystal phases, their level of chemical detail is often not sophisticated enough to attempt a precise quantitative comparison with the experimental observations. Off-lattice CG models, with an explicit representation of the solvent and an *ad hoc* parameterization, offer an excellent compromise between the computational efficiency of lattice models and the complexity of fully atomistic models. They have been applied to describe the self-assembly pathway of proteins,[35] surfactants,[36,37] polymer solutions,[38] and block copolymers.[39] These models are generally benchmarked against structural or dynamical properties obtained from atomistic simulations. This is for instance the approach adopted by Keten and co-workers to investigate the dynamics in melts of methacrylate-based homopolymers, including PMMA.[40]

Despite the practical impact of methacrylate-based copolymers in the formulation of hier-archical porous materials, there are no available CG models to describe their self-assembly. In this work, we propose an off-lattice CG model for PEO-*b*-PMMA and PEO-*b*-PBMA, which can be extended to other methacrylate-based block copolymers in solution. This model can be applied to investigate their self-assembly into complex nanostructures, including bicontinuous and liquid crystal phases, whose characteristic time and length scales are too large for the current computational efficiency of atomistic models. To validate our CG model, we perform extensive Molecular Dynamics (MD) simulations of either single chains dissolved in THF, where inter-chain interactions are virtually absent, or pure polymer systems, where these interactions become dominant, and compare our results with those obtained with an atomistic model. Additionally, to test the transferability of our CG model to different chemical environments, we calculate the monolayer surface pressure-area isotherm of a PEO-*b*-PBMA copolymer at the air-water interface, which is found to be in very good agreement with that measured experimentally.

## 2. Computational Details

All simulations, both atomistic and coarse-grained, were performed by employing the GROMACS 5.0.4 molecular dynamics package.[41]

### 2.1 Atomistic Simulations

We used well-defined classical intra and intermolecular potentials of the GROMOS[42] force-field family to describe interatomic forces of the molecular systems represented as united-atom structures. In particular, parameters of the 54A7 version were adopted.[43] Initial cubic simulation cells with periodic boundaries were constructed by random spatial distribution and rotation of solvent and/or copolymer molecules using the PACKMOL package.[44] The species were initially arranged in the simulation box at relatively low densities to avoid overlap and energy minimization was performed by applying the steepest descent method.

Subsequently, the system was allowed to achieve the equilibrium density in the isothermal-isobaric ($NPT$) ensemble at 300 K and 1 bar. All simulations were performed by integrating the classical equations of motion using the leapfrog MD algorithm[45] with a time step of 2 fs. The temperature was kept constant by the stochastic velocity rescale thermostat[46] with a relaxation time of 0.5 ps. Isotropic pressure coupling was employed using the Berendsen algorithm[47] with a barostat relaxation constant of 3.0 ps and compressibility of $4.5 \times 10^{-5}$ bar$^{-1}$. The cut-off for the non-bonded (dispersion and electrostatic) interactions was set to 1.4 nm excluding standard long-range corrections to the energy and pressure. Finally, a reaction-field approach was applied to handle long-range electrostatic interactions.

## 2.2 Coarse-Grained Simulations

CG simulations were performed in the $NPT$ ensemble, where the equations of motion were numerically integrated using the standard MARTINI time step of 20 fs.[48] The ensemble temperature was set to 300 K using the stochastic velocity rescale thermostat with a relaxation time of 1.0 ps and the pressure was maintained at 1 bar by means of the Berendsen barostat with isotropic pressure coupling characterized with a relaxation constant of 3.0 ps and a compressibility of $5 \times 10^{-5}$ bar$^{-1}$. Three-dimensional periodic boundary conditions were applied. The shift function for dispersion interactions starting from 0.9 nm was employed with a cut-off of 1.2 nm. For systems consisting of copolymer monolayers at the air-water interface, the initial configurations consisted of a water slab bounded by two empty regions that mimic the presence of air by exerting a pressure of 1 bar, whereas the copolymer was arranged in two identical monolayers located at each air-water interface. Each monolayer contained 400 copolymer chains, with their hydrophilic and hydrophobic blocks directed towards water and air, respectively. While air is only implicitly modelled, water is modelled according to the MARTINI force-field (4-to-1 mapping) with no addition of antifreeze particles.[48] Simulations of a duration of 3 $\mu$s were accomplished using the surface tension coupling with the normal pressure to the interface set at 1 bar and leaving the lateral pressure as an independent variable to allow for interface area fluctuations.

## 2.3 Thermodynamic Integration

In the optimization of the monomeric CG model of THF, we select the Lennard-Jones (LJ) parameters for cross-interactions with water and octanol to reproduce the experimental free energy of partitioning between these two fluids. The free energy change upon transferring a THF molecule from octanol to water ($\Delta G_{\text{ow}}$) was approximated as the difference between the free energy of solvation of the molecule in pure octanol ($\Delta G_{\text{o}}^{\text{solv}}$) and pure water ($\Delta G_{\text{w}}^{\text{solv}}$):

$$\Delta G_{\text{ow}} \approx \Delta G_{\text{w}}^{\text{solv}} - \Delta G_{\text{o}}^{\text{solv}} \qquad (1)$$

The individual free energies of solvation were determined by employing the thermodynamic integration (TI) method,[49] in which the Gibbs free energy difference between two macroscopic states is computed by integrating out the ensemble average of the first order derivative of the Hamiltonian of the system, $\mathcal{H}\left(\boldsymbol{q}^N, \boldsymbol{p}^N; \lambda\right)$, with respect to the coupling parameter $\lambda$. In particular, in the configurational part of $\mathcal{H}$, only non-bonded dispersion interactions were coupled to $\lambda$ and the two characteristic states of the system were specified as follows: with $\lambda = 0$, all the interactions were present and with $\lambda = 1$ the molecule did not interact with the solvent. Therefore, assuming reversibility, the energy of solvation of the individual molecule in each solvent was determined by:

$$\Delta G^{\text{solv}} = -\int_{\lambda=0}^{\lambda=1} \left\langle \frac{\partial \mathcal{H}\left(\boldsymbol{q}^N, \boldsymbol{p}^N; \lambda\right)}{\partial \lambda} \right\rangle_{NPT;\lambda} d\lambda \qquad (2)$$

In order to enhance phase space sampling and to better capture variations in the curve $\partial \mathcal{H}\left(\boldsymbol{q}^N, \boldsymbol{p}^N; \lambda\right)/\partial \lambda$ along the pathway connecting the initial and final states of the system, we performed 11 simulations of a duration of 500 ns, corresponding to equally-spaced intermediate values of $\lambda$. The simulations were carried out by inserting a single THF molecule in a cubic box containing either 750 water beads ($\approx$ 3000 real water molecules) or 500 octanol molecules, which were modelled using the MARTINI force-field. All simulations were performed in the $NPT$ ensemble at 300 K and 1 bar. In order to avoid singularities in the potential due to the removal or addition of non-bonded interaction sites, the LJ pair potential was modified into a soft-core interaction.[50]

## 3. Development of the Coarse-Grained Models

Molecular simulations of biological and soft matter systems using classical all-atom force-fields remain computationally challenging. By contrast, CG models provide the necessary efficiency to explore length and time scales of larger orders of magnitude and thus allow for a feasible description of the mesoscale behaviour of a wide spectrum of self-assembling systems.[51] Although generic primitive CG models are able to provide molecular-scale information and trends at a qualitatively level,[52–54] the development of chemical realistic models is key to gain a quantitative insight for a direct comparison with experiments.[55] While deriving CG potential functions with specific parameters for a molecular model, one typically has to identify the targeted properties to measure and integrate out the less relevant degrees of freedom of the higher-resolution reference system. The parameterization of the CG force-field performed here aims to capture thermodynamic, interfacial, and structural properties of real methacrylate-based copolymers. In the design of our CG model, the link with the atomistic representation for the different molecules has been established in line with the MARTINI force-field,[48] which is a versatile model that has been widely used for studying condensed phases of proteins,[56,57] lipids,[58] carbohydrates[59] and polymers.[60–62]

Within the group-contribution MARTINI approach, non-bonded interactions using the spherically symmetrical LJ potential were originally parameterized to allow for the reproduction of densities and free energies of partitioning between water and organic phases of a collection of 18 different resembling chemical groups. In the case of PEO-$b$-PMMA and PEO-*b*-PBMA, we have additionally included structural properties as target for the optimization of the force-field. Probability distributions of bond distances and angles obtained from atomistic simulations are used to derive parameters for the effective bonded interactions. In all the cases, the CG bonded potentials are expressed in terms of simple analytical functions compatible with the MARTINI force-field. The approach presented here provides a systematic means to develop computationally efficient and specific high-level CG potentials incorporating information from both top-down (experimental) and bottom-up (atomistic) approaches.

## 3.1 Single-Segment Coarse-Grained Model for THF

Although THF is a solvent ubiquitously used in a broad spectrum of applications, limited work has been directed to the development of molecular models for the prediction of its thermodynamic properties by direct molecular simulation. Very recently, Garrido *et al.*[63] proposed CG models that allow for an accurate description of the vapour-liquid equilibrium and interfacial properties of THF through molecular dynamics simulations based on the SAFT-$\gamma$ Mie approach,[64] which links macroscopic properties of complex fluids and force-field parameters using a molecular-based equation of state.

In the MARTINI force-field, proposed by Marrink *et al.*,[48] effective non-bonded pair interactions between CG sites are described by a LJ potential, which is truncated at $r_\mathrm{c} = 1.2$ nm and shifted from $r_\mathrm{s} = 0.9$ nm, causing the forces to decay smoothly to zero at $r_\mathrm{c}$. By contrast, the top-down SAFT-$\gamma$ approach uses a Mie (generalized LJ) potential and is based on a high-temperature perturbative expansion, where the long-ranged slowly varying attractive tail of the interaction is crucial to determine the thermodynamic properties of molecular systems. As such, the truncation cut-off radius for MD simulations using the SAFT-$\gamma$ force-field tends to be much larger than that used in MARTINI.[65,66]

We tested the adaptability of the SAFT-$\gamma$ monomeric model of THF to be used in combination with MARTINI molecules by simulating mixtures of non-polarizable water sites and THF in the $NPT$ ensemble. The energy- and length-scale parameters for the Mie potential describing the cross-interactions were determined according to the SAFT-$\gamma$ Mie force-field mixing rules. At all the mixture compositions, water and SAFT-$\gamma$ THF molecules rapidly demixed into two different phases. Notice that the experimental phase diagram of the binary system exhibits miscibility of the liquid phases at the studied thermodynamic conditions.[67] Therefore, the parameterization of a new single-site interaction model of THF fully compatible with the MARTINI force-field was necessary.

Following the MARTINI philosophy, the parameterization of the new CG particle was based on the accurate estimation of the partition free energy between water and octanol liquid phases, whose experimental value is $\Delta G_{\text{ow}} = 2.64$ kJ/mol.[68] In principle, the MARTINI CG sites of non-polar species, $N_a$, could be used to reproduce this feature within a statistical accuracy of $\pm 1$ kJ/mol. Nevertheless, the length- and energy-scale parameters for the self-interaction lead to an underestimation of the bulk density of THF at room conditions. In order to overcome this problem, the segment diameter of the like interaction was increased from the standard value of 0.47 nm to 0.48 nm. We coined this new interaction site containing five heavy atoms as $N_G$. Additionally, the unlike interaction with the water beads was systematically modified to better reproduce the experimental value of $\Delta G_{\text{w}}^{\text{solv}}$ and therefore $\Delta G_{\text{ow}}$. Cross-interactions between THF and octanol were not altered. The free energy calculations were performed on the basis of the TI technique described in section 2.3.

The vapour-liquid phase equilibrium envelope and interfacial properties of the new CG model of THF were also determined by direct coexistence simulations in the canonical ($NVT$) ensemble. In particular, we first equilibrated a liquid phase of $N = 12500$ THF beads in the $NPT$ ensemble. After equilibration, the simulation box was expanded to three times its original size along the $z$ direction, and the system was allowed to achieve, in the $NVT$ ensemble, a new equilibrium state, where two liquid/vapour interfaces were formed. The properties of these coexisting phases were then measured at different temperatures by averaging the simulation trajectories over the last 300 ns.

The diagonal components of the pressure tensor, $\boldsymbol{P}$, were used in order to estimate the vapour-liquid interfacial tension, $\gamma$, and the equilibrium vapour pressure, $P^{\text{vap}}$. While the latter corresponds to the normal component of $\boldsymbol{P}$, i.e. $P^{\text{vap}} = P_{zz}$, the former was calculated using the normal and tangential components:

$$\gamma = \frac{1}{2} L_z \left[ P_{zz} - \frac{1}{2} \left( P_{xx} + P_{yy} \right) \right], \qquad (3)$$

where $L_z$ is the box dimension in the $z$ direction and the factor $1/2$ is a correction for the presence of the two interfaces. To estimate the critical temperature, $T_c$, we fitted the results to the scaling law for the density[69] given by:

$$\rho_l - \rho_v = A_0 \left( T - T_c \right)^\beta \qquad (4)$$

and the law of rectilinear diameters

$$\frac{1}{2} \left( \rho_l + \rho_v \right) = \rho_c + B_0 \left( T - T_c \right), \qquad (5)$$

where $\rho_l$ and $\rho_v$ are, respectively, the density of the liquid and vapour phase, $A_0$ and $B_0$ are fitting parameters, and $\beta = 0.32$ the three-dimensional Ising critical exponent.[70]

## 3.2 Coarse-Grained Models for PEO-*b*-PMMA and PEO-*b*-PBMA

The low molecular weight diblock copolymers PEO$_6$-*b*-PMMA$_7$ and PEO$_6$-*b*-PBMA$_7$ were used as reference systems to build our CG model. Experimentally, this type of copolymers are usually synthesized via atom transfer radical polymerization (ATRP),[71,72] which causes the final molecular structures to have a residual group from the macroinitiator linking the two blocks. In this work, for the sake of simplicity, the methacrylate and PEO blocks are as-sumed to be directly bonded in both copolymers. In the following, we present the two stages for the development of our MARTINI-based CG model: (*i*) mapping of the high-resolution atomistic structure to the CG representation along with the definition of non-bonded interactions, and (*ii*) derivation of intramolecular (bonded) potentials.

### 3.2.1 Mapping and Non-Bonded Interactions

Although several efforts have been recently made to define automated mapping procedures to transform detailed atomistic configurations into CG structures,[73–75] a formal statistical mechanical analysis indicates that a unique translation between both levels of representation is not possible.[76] In this work, the selection of the mapping strategy is limited by the number of CG sites present in the MARTINI parameterization. MARTINI beads with a LJ length scale parameter $\sigma = 0.47$ nm typically represent a 4-to-1 mapping, whereas smaller beads with $\sigma = 0.43$ nm have been used to represent 3-to-1 and 2-to-1 mappings (for a complete description of the different MARTINI CG sites see the original reference[48]).

Our mapping scheme is schematically sketched in figure 1, where we show the CG representation of the species studied in this work, indicated by transparent beads, along with its atomistic counterpart, represented by solid segments in the background. The hydrophilic PEO block consists of one terminal and six bridging beads, mimicking, respectively, the ending $HO-CH_2-$ group and the repeating $-CH_2-O-CH_2-$ monomers. The former is described by an $SP_h$ site, while the latter by an $SP_0$ site, in accordance with the CG MARTINI model of PEO proposed by Rossi *et al.*[77] The repeating units of the hydrophobic PMMA and PBMA blocks display similar molecular structures as they only differ in the length of the aliphatic side-chain group attached to the methacrylate group. In particular, both methacrylate blocks incorporate seven backbone beads, each mimicking a group of three carbon atoms and here referred to as $SC_{1,B}$, in agreement with the original MARTINI nomenclature.[48] The subscript B, for backbone, is employed to distinguish this bead from the terminal bead in the side chain of the PBMA block, referred to as $SC_{1,T}$, but the LJ parameters to describe their non-bonded interactions are identical. Additionally, the side chains of both PMMA and PBMA also incoporate an ester group, denoted as $N_a$, which occupies a terminal position in the former and a bridging position in the latter block. Finally, in the same figure, we also show the CG representation of a THF molecule, referred to as $N_G$, which incorporates four carbon and one oxygen atoms. In the case of the copolymers, the energy- and length-scale parameters for the interaction between pairs of LJ particles were not altered in order to retain the interfacial properties of the original CG MARTINI optimization.

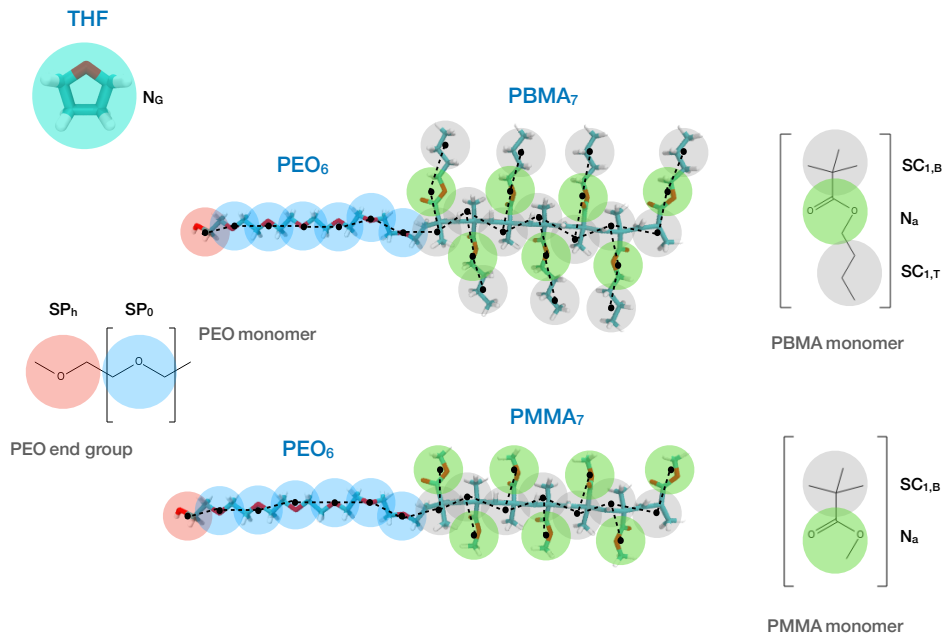

Figure 1: Coarse-grained (transparent beads) and atomistic (solid segments in the background) representations of the species investigated in this work: THF, $PEO_6$-$b$-$PMMA_7$ and $PEO_6$-$b$-$PBMA_7$. PEO, PBMA and PMMA blocks are highlighted separately to better identify their repeating units. In particular, $SP_h$ and $SP_0$ indicate, respectively, the terminal and bridging beads of the PEO block; and $SC_{1,T}$, $N_a$ and $SC_{1,B}$ the terminal, bridging, and backbone beads of the PBMA block. The last two beads are also part of the PMMA block, whose side chains only consist of one $N_a$ bead. THF is modelled by a single bead, here defined as $N_G$. The black dashed lines in the polymer chains indicate the connectivity between CG sites. Red, white and light blue solid segments in the atomistic model represent, respectively, oxygen, hydrogen and carbon atoms. See text for additional details.

### 3.2.2 Bonded Interaction Optimization

The intramolecular interaction potentials of $PEO_6$-$b$-$PMMA_7$ and $PEO_6$-$b$-$PBMA_7$ have been derived by applying the Boltzmann Inversion (BI)[78] of probability distributions of bond distances and angles obtained from atomistic simulations. However, unlike earlier BI-based parameterizations that employed a single reference system,[38,60] here we define our target distributions as the average probability of bonded degrees of freedom in ($i$) dilute polymer solutions of THF and ($ii$) pure polymer systems. This approach is especially beneficial to capture the most significant features of both environments and thus enhances the applicability of our CG model, in line with the framework of multistate force-field optimization.[40,55]

In order to generate the atomistic target probability distributions, extensive atomistic simulations consisting of either single chains dissolved in THF (10,000 solvent molecules) or pure polymer systems (consisting of 300 chains) were performed using the GROMOS force-field. The resulting trajectories were converted into the CG representation by mapping the centres of mass (COMs) of groups of constituent atoms according to the approach illustrated in figure 1. From the mapped trajectories, probability distributions of bond distances and angles were computed for each copolymer in each environment. Using BI, the initial average potentials for every type of bond and angle were estimated as follows

$$\overline{U}_{\text{bond}}^{\text{initial}}(b) = -\frac{1}{2}k_BT\left[\ln\left(P_{\text{sol}}(b)\right) + \ln\left(P_{\text{pol}}(b)\right)\right] \qquad (6)$$

$$\overline{U}_{\text{angle}}^{\text{initial}}(\theta) = -\frac{1}{2}k_BT\left[\ln\left(P_{\text{sol}}(\theta)\right) + \ln\left(P_{\text{pol}}(\theta)\right)\right] \qquad (7)$$

where $P_{\text{sol}}$ and $P_{\text{pol}}$ are the mapped probability distributions obtained in solution and pure polymer systems, respectively. We assume that the choice of the average distributions as target for the CG potential derivation should in principle enhance the transferability of the models across different simulation systems involving the two different chemical environments.

Since the initial potentials derived with equations 6 and 7 implicitly incorporate many-body effects, they can be considered as potentials of mean force, thus, CG simulations using these potentials did not reproduce the atomistic target distributions and therefore a refinement procedure was implemented to correct the expressions into effective potentials. In order to keep the model compatible with the MARTINI force-field, the initial inverted probability distributions for the different bonded interactions were fitted to the following classical harmonic potential functions

$$\overline{U}_{\text{bond}}(b) = \frac{1}{2}k_{\text{b}}\left(b - b_0\right)^2 \qquad (8)$$

$$\overline{U}_{\text{angle}}(\theta) = \frac{1}{2}k_\theta \left(\cos(\theta) - \cos(\theta_0)\right)^2 \qquad (9)$$

The characteristic force constants and equilibrium bond lengths and angles were then systematically tuned until a reasonable agreement between atomistic and CG distributions was achieved.

## 4. Results and discussion

### 4.1 Thermodynamic and Interfacial Properties of CG THF

In this section, we present the simulation results obtained with our monomeric CG model of THF. We assess and validate the performance of the parameterized model in reproducing the experimental bulk density and free energy of partition between water and octanol in the liquid state. As a secondary test, we examine the vapour-liquid interfacial properties and compare our results with literature simulation data obtained with other more sophisticated molecular models. The calculated average equilibrium bulk density of our CG model of THF at 1 bar and 300 K matches to a high accuracy the experimental value. Likewise, the potential optimization for describing cross-interactions with other CG particles based on TI calculations allow for a close reproduction of the free energy of partition between water and octanol. The optimized LJ parameters for the like and unlike interactions are reported in table 1, while in table 2 the thermodynamic data obtained with our model are summarized and compared with experimental results.[67,68]

The vapour-liquid phase equilibrium of the CG model was studied by direct coexistence simulations in the $NVT$ ensemble. In order to characterize the liquid and vapour phases in equilibrium at a given temperature, density profiles ($\rho(z)$) along the $z$ direction perpendicular to the interface were determined by discretizing the simulation box in 250 slabs. The bulk liquid and vapour densities were then estimated by averaging $\rho(z)$ away from the interface. The VLE envelope for the CG model presented in this work is depicted in figure 2. Coexistence curves obtained with other molecular models reported in the work by Garrido *et al.*[63] are also included for comparison. According to the results presented by the authors, the SAFT-$\gamma$ models were able to accurately reproduce the experimental coexistence densities. As can be observed, our model is able to closely predict the correct temperature-density phase diagram, although it slightly overestimates the actual vapour and liquid densities in the whole range of temperatures. In particular, positive deviations in the predicted fluid density become more important in the vapour branch as the temperature increases. Such a discrepancy has been identified as one of the limitations of the MARTINI force-field,[48] in which the vaporization free energies of different compounds although well approximated, do not match quantitatively the real values due to the low stability of the condensed fluid with respect to the vapour phase. This effect can also be noticed in the vapour pressure ($P^{\text{vap}}$) and surface tension ($\gamma$) curves illustrated in figure 3.

Table 1: CG force-field parameters for THF. $P_4$, $P_1$ and $C_1$ denote the standard MARTINI CG sites used for water ($P_4$) and octanol ($C_1$ and $P_1$), whereas $N_G$ corresponds to the bead type parameterized in this work to model THF.

| Pair | $\epsilon_{ij}$ (kJ/mol) | $\sigma_{ij}$ (nm) |
|---|---|---|
| $N_G$-$N_G$ | 4.000 | 0.480 |
| $N_G$-$P_4$ | 4.120 | 0.470 |
| $N_G$-$P_1$ | 4.500 | 0.470 |
| $N_G$-$C_1$ | 2.700 | 0.470 |

Table 2: Thermodynamic Properties of THF$^a$.

| Property | CG | Experimental |
|---|---|---|
| $\Delta G_{\text{ow}}$ (kJ/mol) | 2.67 ± 0.016 | 2.64[68] |
| $\rho_{\text{bulk}}$ (kg/m$^3$) | 896.0 | 882.1[67] |

$^a$ Experiments have been performed in the range 298-300 K and simulations at 300 K.

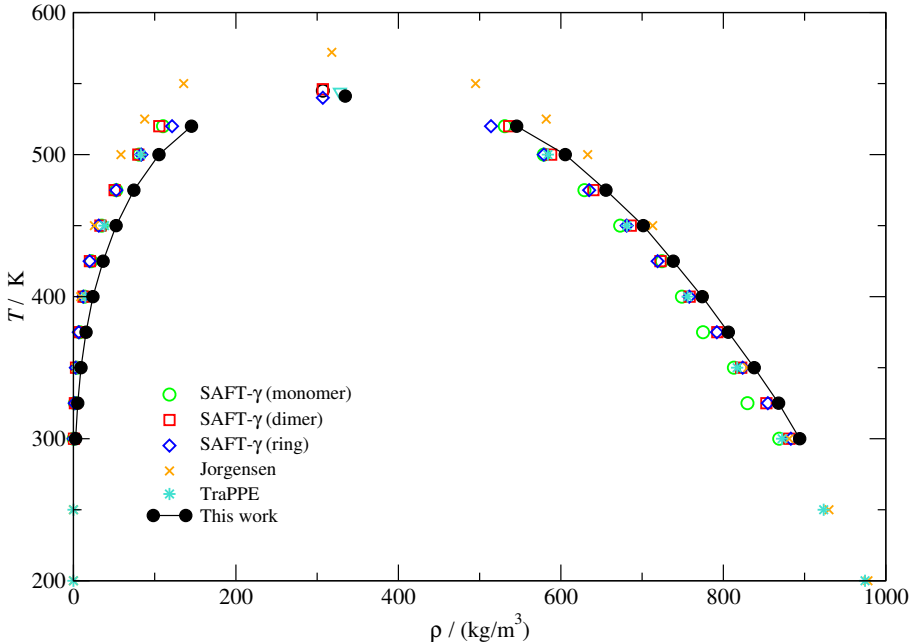

Figure 2: Temperature-density vapour-liquid coexistence envelope for THF. Data points indicated for the SAFT-$\gamma$ (monomer), SAFT-$\gamma$ (dimer), SAFT-$\gamma$ (ring), Jorgensen and TraPPE models are compiled from ref.[63] The continuous black line is added as a guide to the eye.

The lack of a full agreement between our results and those displayed in figures 2 and 3 can also be due to the substantial sensitivity of the interfacial properties to the potential truncation.[63,79] As previously mentioned, we have used the standard MARTINI truncation procedure, which restricts the calculation of dispersion interactions up to about $2.5\sigma$ without real-space long-range corrections. However, it has been established that a cut-off larger than six diameters is necessary to provide a reliable description of interfacial properties.[79] Despite these limitations, our model describes very well the thermodynamic properties of the condensed phase ($\Delta G_{\text{ow}}$ and $\rho_{\text{bulk}}$) at the state point selected to parameterize the CG potentials for the copolymers and no further calibration was performed.

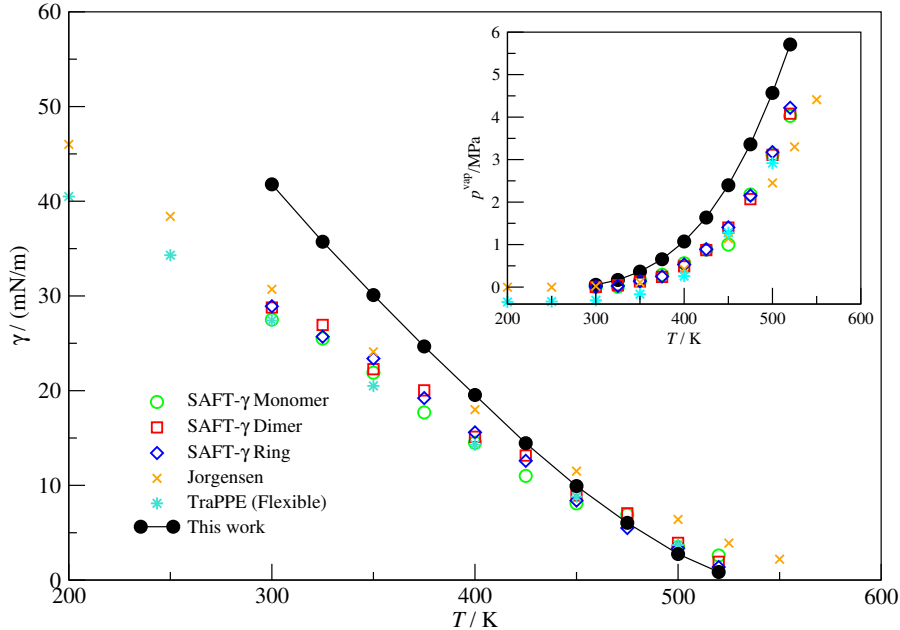

Figure 3: Surface tension and equilibrium vapour pressure (inset plot) as a function of temperature for THF. Data points indicated for the SAFT-$\gamma$ (monomer), SAFT-$\gamma$ (dimer), SAFT-$\gamma$ (ring), Jorgensen and TraPPE models are compiled from ref.[63] The continuous black line is added as a guide to the eye.

## 4.2 Coarse-Grained Potentials for Copolymers

A major limitation in coarse-graining is the lack of thermodynamic consistency between various representation levels.[80] Due to the presence of many-body effects, which are implicitly incorporated into simple CG potentials, the derived parameters tend to be generally temperature, composition and density dependent. Thus, in principle, for the two studied systems, a pure polymer and a dilute solution, two different parameterizations should be performed. In the approach adopted here, we instead have defined the target distributions as the average probability density of bonded conformations across the two environments in pursuit of capturing the foremost features of each of them. In the following, we check the validity of this assumption for the two methacrylate-based copolymers. In order to retain the interfacial properties of the original MARTINI optimization, the energy- and length-scale parameters for the intermolecular pair interactions of the chain beads were not altered. The complete list of parameters for all the non-bonded interaction between CG beads, including the new model for THF, is given in table 3.

Table 3: Force-field parameters for LJ interactions between CG sites. $SC_{1,i}$, with i = B or T, and $N_a$ correspond to the original MARTINI CG sites,[48] whereas $SP_h$ and $SP_0$ to the PEO sites proposed by Rossi et al.[77] $N_G$ is the bead type parameterized in this work to model THF.

| Pair | $\epsilon_{ij}$ (kJ/mol) | $\sigma_{ij}$ (nm) | Pair | $\epsilon_{ij}$ (kJ/mol) | $\sigma_{ij}$ (nm) |
|---|---|---|---|---|---|
| $N_G$-$N_G$ | 4.000 | 0.480 | $SP_h$-$SP_0$ | 3.375 | 0.430 |
| $SP_h$-$SP_h$ | 3.750 | 0.430 | $SP_h$-$SC_{1,i}$ | 2.025 | 0.430 |
| $SP_0$-$SP_0$ | 3.375 | 0.430 | $SP_h$-$N_a$ | 4.500 | 0.470 |
| $SC_{1,i}$-$SC_{1,i}$ | 2.625 | 0.430 | | | |
| $N_a$-$N_a$ | 4.000 | 0.470 | $SP_0$-$SC_{1,i}$ | 2.025 | 0.430 |
| | | | $SP_0$-$N_a$ | 4.500 | 0.470 |
| $N_G$-$SP_h$ | 4.500 | 0.470 | | | |
| $N_G$-$SP_0$ | 4.500 | 0.470 | $SC_{1,i}$-$N_a$ | 2.700 | 0.470 |
| $N_G$-$SC_{1,i}$ | 2.700 | 0.470 | | | |
| $N_G$-$N_a$ | 4.500 | 0.470 | | | |

The LJ parameters describing the non-bonded interactions of the copolymers' methacrylate blocks have been set to reproduce, within a reasonably good approximation, the experimental densities of their repeating units, namely methylmetacrylate (MMA) and butylmethacrylate (BMA). In Table 4, we report the MMA and BMA densities as measured experimentally[81] along with those estimated by simulation using our CG model. Differences in the order of $\approx 6\%$ are comparable to those obtained in the MARTINI-based parameterization of the PEO block by Lee et al.[38]

Table 4: Simulated and experimental densities of PMMA and PBMA monomers[a].

| Monomer | CG (kg/m$^3$) | Experimental (kg/m$^3$) |
|---|---|---|
| MMA | 996.300 | 933.700[81] |
| BMA | 948.200 | 893.600[81] |

[a] Experiments have been performed between 298 K and 300 K, while simulations at 300 K.

### 4.2.1 Bond Angle and Length Probability Distributions

In this section, we present and analyze the bond angle and length probability distributions derived for $PEO_6$-$b$-$PBMA_7$ and refer the reader to the supporting information for similar results on $PEO_6$-$b$-$PMMA_7$. For both methacrylate-based diblock copolymers, the parameterization of the bond stretching potentials was considered first and, subsequently, that of the angle potentials. We report in figure 4 the distributions of bond lengths between all the possible pairs of bonded sites as obtained from CG simulations (empty symbols) and compare them with the atomistic simulation results (solid and dashed lines) obtained in dilute polymer solutions and pure polymer systems. Additionally, we also include the resulting effective potential obtained from Eq. 6. As a general tendency, the probability distributions from the atomistic simulations of the pure polymer and polymer solution do not differ significantly from each other, being the most relevant deviation measured for the $SC_{1,B}$–$SC_{1,B}$ bond length, which is peaked at two different values of $b$, that is approximately 0.29 nm in solution and 0.32 nm in the pure polymer. Although most of the profiles closely follow a Gaussian distribution, a left-skewed probability distribution is observed for $SP_h$–$SP_0$ and $SP_0$–$SP_0$ bonds, unveiling the occurrence of a larger number of accessible configurations at relatively short distances. In all the cases, however, the average first Boltzmann-inverted CG potentials was good enough to capture the mean equilibrium values of the atomistic distributions and was then directly fitted to the classical harmonic function of Eq. 8. Only a minor tuning of the force constants was needed to better reproduce the width of the distribution functions.

Similarly, for the derivation of the average angle bending potentials, the direct Boltzmann-inverted atomistic distributions (Eq. 7) were taken as initial guess and fitted to the potential function given by Eq. 9. The equilibrium angles and force constants were systematically modified in order to reproduce the average probability distributions of the atomistic systems. In figure 5, we present these distributions along with the resulting effective potentials. The angles between segments $SC_{1,B}$–$SC_{1,B}$–$N_a$ and $SP_0$–$SC_{1,B}$–$N_a$ (see figure 1) were not included in the parameterization, because their distributions were defined by the remaining angle types. To limit the complexity of the problem, we decided to neglect the effect of the dihedral angles at the CG level. In the case of the methacrylate block of the copolymers, the computed atomistic distributions for the backbone beads ($SC_{1,B}$) are comparable to those obtained in other previous PMMA CG parameterizations,[40,82,83] where distinct atomistic force-fields have been used and different mappings have been adopted. In particular, the width of the angle distribution with non-zero probability extending from $\theta \approx 80°$ to $\theta \approx 160°$ and with the maximum located at $\theta \approx 120°$, agrees with the results by Keten and co-workers,[40] which were obtained by employing the general-purpose Dreiding force-field. The bond distance distribution for $SC_{1,B}$–$SC_{1,B}$ is however slightly shifted to larger distances, with the mean value at about $b = 0.30$ nm instead of $b = 0.28$ nm. This small difference reflects the different mapping strategy adopted by the other authors, in which, the backbone bead center is located and fixed at the quaternary carbon atom in the methacrylate group, whereas in our case, the bead center, calculated as the center of mass of the group with three carbon atoms, can fluctuate as explained above.

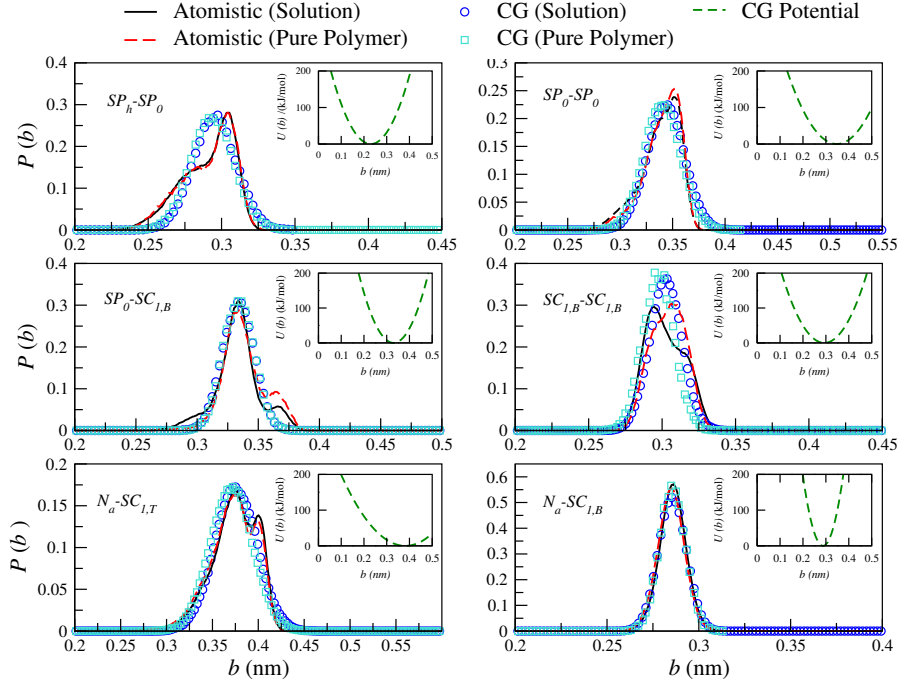

Figure 4: Probability distributions of bond lengths ($b$) between the different CG sites of a $PEO_6$-$b$-$PBMA_7$ chain. Distributions obtained from atomistic simulations of single chains in THF (black solid curves) and pure polymer systems (red dashed curves) are included along with the CG distributions (empty symbols). In the insets, the corresponding effective potentials as calculated from Eq. 6 are reported.

As observed for the bond length distributions, the angle distributions calculated in a polymer solution does not significantly differ from that obtained in a pure polymer. Interestingly, the main discrepancy is found for the angle formed by the beads across the hydrophilic and hydrophobic blocks, that is $SP_0$–$SC_{1,B}$–$SC_{1,B}$. The secondary peak found in solution at approximately $\theta = 150°$ is not observed in a pure polymer system, where this probability distribution appears to be restricted to smaller angles. As a result, our CG model is not able to simultaneously reproduce both distributions and clearly misses this secondary peak in solution. The remaining CG mapping shows a reasonably good agreement, although the occurrence of configurations with relatively small angles is slightly underestimated.

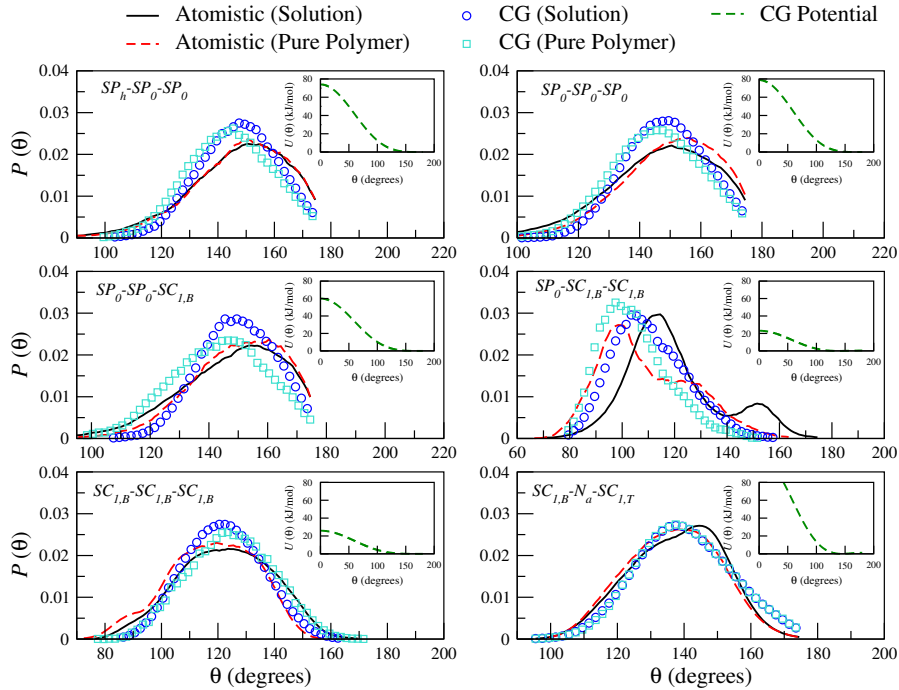

Figure 5: Probability distributions of angles ($\theta$) between the different CG sites comprising a $PEO_6$-$b$-$PBMA_7$ chain. Distributions obtained from atomistic simulations of single chains in THF (black solid curves) and pure polymer systems (red dashed curves) are included along with the CG distributions (empty symbols). In the insets, the corresponding effective potentials as calculated from Eq. 7 are reported.

In general, the reference atomistic distributions for bond lengths and angles are similar regardless of the chemical environment and the average behaviour can be adequately reproduced with the unique parameterized CG analytical potential functions. We additionally observe that the derived set of intramolecular parameters can be applied to model the copolymer $PEO_6$-$b$-$PMMA_7$ (see suporting information) as well as methacrylate-based copolymers of different architecture, as we will show in the follwing sections. The full set of parameters employed to describe the bonded interactions are reported in table 5.

## 4.3 Structural Properties

To test the extent of validity of our CG model, we estimate the size and chain conformation of a number of methacrylate-based copolymers, whose hydrophilic block length is kept constant to 6 beads, whereas the methacrylate block length is varied between 3 and 18 beads. To this end, we calculate the radius of gyration, $R_\text{g}$, and compare the results with atomistic simulations in THF solution and in pure polymer systems at 300 K. In particular, the radius of gyration is calculated as

$$R_\text{g} = \sqrt{\frac{1}{N} \sum_{i=1}^{N} \langle (\boldsymbol{r}_i - \boldsymbol{r}_\text{cm})^2 \rangle}, \qquad (10)$$

where $N$ is the total number of beads in a copolymer chain, whereas $\boldsymbol{r}_i$ and $\boldsymbol{r}_\text{cm}$ are the position vectors of a generic bead $i$ and the chain center of mass, respectively. In table 6, we report the values of $R_\text{g}$ for a family of PEO$_6$-$b$-PBMA$_m$ and PEO$_6$-$b$-PBMA$_m$ copolymers, with $m = (3, 7, 14, 18)$.

Table 5: Force-field parameters for the intramolecular interactions of PEO$_6$-$b$-PMMA$_7$ and PEO$_6$-$b$-PBMA$_7$ copolymers in solutions and pure polymer systems.

| Bonded Interaction | $b_0$ (nm) / $\theta_0$ (deg) | $k_\text{b}$ (kJ mol$^{-1}$ nm$^{-2}$) / $k_\theta$ (kJ mol$^{-1}$) |
|---|---|---|
| SP$_\text{h}$-SP$_0$ | 0.299 | 12500 |
| SP$_0$-SP$_0$ | 0.348 | 8500 |
| SP$_0$-SC$_{1,\text{B}}$ | 0.335 | 16300 |
| SC$_{1,\text{B}}$-SC$_{1,\text{B}}$ | 0.289 | 21100 |
| SC$_{1,\text{B}}$-N$_\text{a}$ | 0.282 | 17000 |
| SC$_{1,\text{T}}$-N$_\text{a}$ | 0.383 | 5000 |
| SP$_\text{h}$-SP$_0$-SP$_0$ | 180 | 37 |
| SP$_0$-SP$_0$-SP$_0$ | 170 | 40 |
| SP$_0$-SP$_0$-SC$_{1,\text{B}}$ | 180 | 30 |
| SP$_0$-SC$_{1,\text{B}}$-SC$_{1,\text{B}}$ | 139 | 15 |
| SC$_{1,\text{B}}$-SC$_{1,\text{B}}$-SC$_{1,\text{B}}$ | 175 | 13 |
| SC$_{1,\text{B}}$-N$_\text{a}$-SC$_{1,\text{T}}$ | 144 | 67 |

For all the systems investigated, the agreement between CG and atomistic simulations is very good. In general, we notice that our CG model tends to overestimate the size of smaller chain and, by contrast, underestimates the size of larger chains, both in solution and in the pure polymer. Nevertheless, this tendency is very light and often within the estimated error.

Table 6: Radius of gyration ($R_\mathrm{g}$) for atomistic and CG methacrylate-based copolymers in pure polymer and THF solution.

|  | Solution | | Pure Polymer | |
| --- | --- | --- | --- | --- |
| $PEO_6$-$b$- | CG (nm) | Atomistic (nm) | CG (nm) | Atomistic (nm) |
| $MMA_3$ | 0.787 ± 0.081 | 0.743 ± 0.088 | 0.781 ± 0.001 | 0.760 ± 0.003 |
| $MMA_7$ | 0.841 ± 0.085 | 0.800 ± 0.109 | 0.851 ± 0.001 | 0.829 ± 0.001 |
| $MMA_{14}$ | 0.939 ± 0.088 | 0.988 ± 0.106 | 0.908 ± 0.001 | 0.937 ± 0.004 |
| $MMA_{18}$ | 1.050 ± 0.117 | 1.060 ± 0.158 | 0.970 ± 0.001 | 1.012 ± 0.001 |
| $BMA_3$ | 0.829 ± 0.075 | 0.787 ± 0.090 | 0.821 ± 0.001 | 0.784 ± 0.002 |
| $BMA_7$ | 0.884 ± 0.062 | 0.835 ± 0.081 | 0.864 ± 0.002 | 0.824 ± 0.001 |
| $BMA_{14}$ | 0.991 ± 0.068 | 1.013 ± 0.093 | 0.970 ± 0.001 | 0.966 ± 0.004 |
| $BMA_{18}$ | 1.070 ± 0.090 | 1.092 ± 0.091 | 1.072 ± 0.001 | 1.082 ± 0.002 |

As a matter of fact, the associated standard deviations show that the distribution of the copolymer's chain conformation overlap, suggesting that the derived CG potentials allow for a sampling of the phase space configurations accesible to the atomistic models. Substantial differences in atomistic chain conformation in the dilute solution and pure polymer systems are not observed and the mean values of the radius of gyration are reproduced with the unique set of CG parameters, showing the universality and transferability of the potentials to work under the constraints imposed by the two different chemical environments.

### 4.3.1 Temperature Dependence of the Model

It is well-established that, in the development of a CG model, the drastic reduction of the degrees of freedom has a dramatic impact on the thermodynamic properties of the system and especially affects the balance between enthalpy and entropy.[84] More specifically, the reduction of the entropic terms is somehow compensated by a reduction in the enthalpic terms of the model in such a way that the system free energy is maintained more or less constant and similar to that calculated by atomistic models. However, modifying the enthalpy can have a remarkable influence on the temperature dependence of the CG model, which, in principle, might not be correct. Consequently, one would not expect a CG potential transferable over thermodynamic states different from that at which the parameterization was performed. Therefore, to gauge the range of applicability of our CG model, the radius of gyration of PEO$_6$-$b$-PBMA$_7$ in solution was computed at temperatures between 300 K and 330 K and then benchmarked against the results from atomistic calculations. We restrict our analysis to these temperatures because it is in this range that the synthesis of polymeric self-assembled structures in THF, whose boiling temperature at atmospheric pressure is approximately 339 K, is usually performed.[72] The data reported in table 7 show a very weak dependence of $R_g$ on $T$, at least in this range of temperatures.

Table 7: Radius of gyration ($R_g$) for atomistic and CG PEO$_6$-$b$-PBMA$_7$ in THF solution at different temperatures.

| Temperature (K) | Solution | |
| --- | --- | --- |
|  | CG (nm) | Atomistic (nm) |
| 300 | 0.884 ± 0.062 | 0.835 ± 0.081 |
| 310 | 0.883 ± 0.059 | 0.829 ± 0.078 |
| 320 | 0.886 ± 0.070 | 0.838 ± 0.080 |
| 330 | 0.886 ± 0.068 | 0.833 ± 0.076 |

### 4.4 Copolymer Monolayer Adsorption at the Air-Water Interface

Due to their amphiphilic nature, PEO-$b$-PBMA copolymers are interfacially active molecules and can act as compatibilizing agents for two-phase systems, such as air-water systems. At relatively low concentrations, most of the molecules tend to migrate from the bulk to the interface, with the PEO block preferentially in the liquid phase and the PBMA block trying to minimize its contact with water and thus pointing towards the gas phase. At increasing concentration, but still below the CMC, a monolayer is formed. Monolayers of several PEO-$b$-PBMA copolymers have been observed experimentally and characterized by their Langmuir pressure-area ($\Pi - A_L$) isotherms as well as by X-ray reflectivity (XR) measurements.[85]

In general, it has been observed that, at a fixed temperature, non-fluorinated copolymers exhibit a succession of phase transitions, from gas to solid, as the area per molecule is decreased. At low to moderate densities, the monolayers are found in the so-called liquid-expanded (LE) state,[86,87] where the PEO blocks behave like random coils in solution (pancake configuration).[85] By contrast, at larger densities, the system is in a liquid-condensed (LC) phase and the polymer enters the "brush" regime, where the individual chains tend to reorient and adopt stretched conformations. For copolymers of different PBMA/PEO block-length ratio, defined as $f = N_{\text{BMA}}/N_{\text{EO}}$, where $N_{\text{BMA}}$ and $N_{\text{EO}}$ are, respectively, the number of repeating units in the PBMA and PEO blocks, Li *et al.* obtained Langmuir isotherms with similar features, and well-defined phases and phase transitions.[85] A full insight into the structure of the copolymer chains within the self-assembled monolayer can, however, only be achieved by molecular simulation.

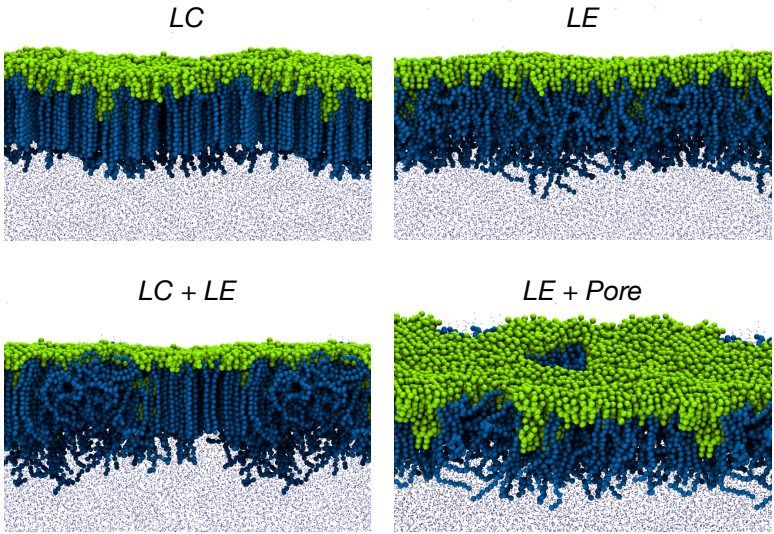

Figure 6: Typical configurations of equilibrated $PEO_{15}$-$b$-$PBMA_5$ monolayers containing 400 chains at the air-water interface and forming liquid-condensed (LC), liquid-expanded (LE), coexisting LC + LE, and perforated LE phases. The PEO segment is shown in dark blue, whereas the PBMA block corresponds to the green-coloured beads. The small blue points represent water beads.

To this end, we performed CG simulations at different values of the monolayer surface tension ($\gamma_{\text{m}}$) in water-air systems containing $PEO_{15}$-$b$-$PBMA_5$ at 300 K. The block-length ratio of this copolymer, $f \approx 0.3$, is similar to that of $PEO_{113}$-$b$-$PBMA_{30}$ employed in the experimental calculation of the Langmuir adsorption isotherms.[85] A number of configurations obtained from these simulations are shown in figure 6, where we show the LE and LC phases that have also been observed experimentally.[85] In particular, the transition from the LE phase to the coexistence LE+LC region and then to the LC phase is promoted by an increasing lateral compression exerted on the monolayer at constant normal pressure.

To compare our results with experiments, we notice that in PEO-based copolymers adsorbed at the air-water interface, the surface area per molecule has been found to increase with the number of monomers, $N_{\text{EO}}$, in the PEO block according to the scaling law $A_\text{L} \approx N_{\text{EO}}^{1/2}$.[88,89] By applying this scaling law, we can rescale the results obtained experimentally by Li to the case of a smaller copolymer with the same block-length ratio and $N_{\text{EO}} = 15$. In particular:

$$A_\text{L}(N_{\text{EO}}) = A_\text{L}(N_{\text{EO}}^*) \left(\frac{N_{\text{EO}}}{N_{\text{EO}}^*}\right)^{1/2}, \qquad (11)$$

where the superscript $^*$ refers to the actual experimental system, where $N_{\text{EO}} = 113$. In figure 7, we show the dependence of the monolayer surface tension on the rescaled molecular area as obtained experimentally from the standard relation $\gamma_\text{m}^{\text{EXP}} = \gamma_\text{aw}^{\text{EXP}} - \Pi^{\text{EXP}}$, where the experimental value of the surface tension at the air-water interface, $\gamma_\text{aw}^{\text{EXP}} = 72$ mN/m, was used. The solid points refer to our simulation results. In particular, we applied surface tension values from 2 to 40 mN/m. At each point, the corresponding area per molecule ($A_\text{L}$) at equilibrium was obtained by dividing the cross-sectional area of the simulation box by the total number of copolymer chains in the monolayer. For molecular areas, smaller than $A_\text{L} = 0.57$ nm$^2$, our simulation results indicate the formation of an LC phase. At $A_\text{L} = 0.57$ nm$^2$, LC and LE phases coexist. Finally, for intermediate values of areas per molecule the system exhibit an LE phase, which transforms into a perforated LE phase at the lower densities.

As it can be observed, the CG model qualitatively predicts the correct trend of the monolayer surface tension as a function of the molecular area for values of $A_L$ above 0.65 nm$^2$, but it deviates significantly at high densities, where the system exhibits an LC phase. Additionally, the slope of our calculated isotherm appears to be steeper in comparison to that of the experimental curve, suggesting that our model would overestimate the area compressibility modulus of the monolayer. Based on the aforementioned observations, the model seems to perform well for describing the interfacial properties of the real system in the LE region, however, due to the small number of calculated state points, a definitive conclusion cannot be stated and further investigation on the potentialities of the model to enable the identification of properties not accessible from experiments remains open.

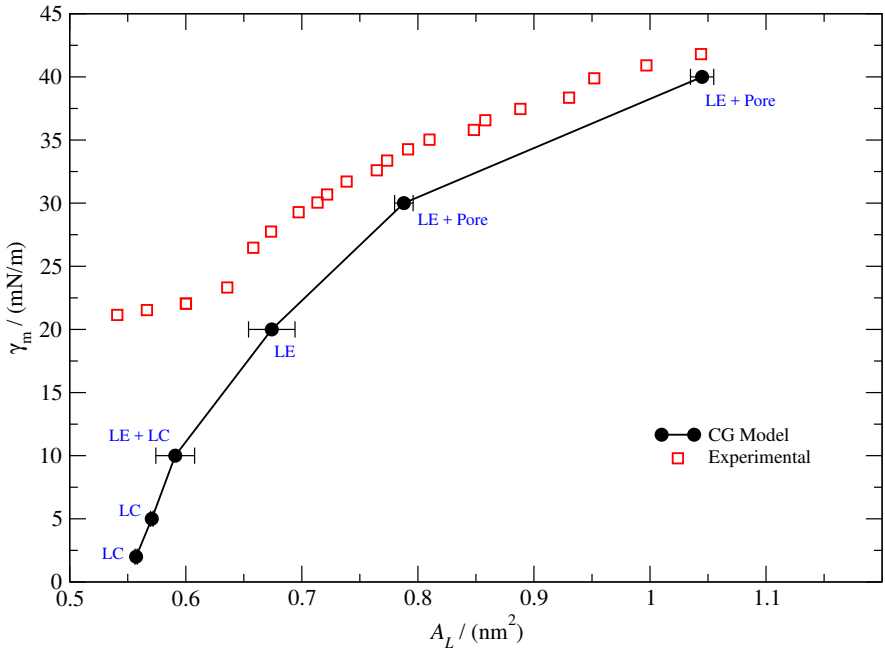

Figure 7: Monolayer surface tension−molecular area isotherm at 300 K. The solid black dots correspond to our simulation results for a PEO$_{15}$-$b$-PBMA$_5$ monolayer ($f \approx 0.3$). The solid line, which is a guide for the eye, exhibits perforated LE, LE, LC and coexisting LC and LE phases. The red empty squares correspond to the experimental data points measured for a PEO$_{113}$-$b$-PBMA$_{15}$ monolayer ($f \approx 0.3$) and rescaled according to Eq. 11 (adapted and reprinted with permission from [J. Phys. Chem. B 113, 35, 11841-11847]. Copyright 2009 American Chemical Society).

# 5. Conclusions

In summary, we have designed and discussed a CG potential derivation for THF and two methacrylate-based copolymers, namely, PEO-*b*-PMMA and PEO-*b*-PBMA using a hybrid structural-thermodynamic approach within the framework of the MARTINI force-field. To represent THF as a monomeric fluid, we introduced a new bead type, coined as $N_G$, which is able to reproduce the experimental values of the bulk density and free energy of partitioning between water and octanol in the liquid state. To enhance the transferability of the intramolecular CG potentials and capture the distinguishing features of different chemical environments, our copolymer models were successfully parameterized by matching the average probability distributions of the chain conformation in very dilute THF solutions and in pure polymer systems as obtained from atomistic simulations.

Our multi-phase parameterization has proven to perform well in describing PEO-*b*-PMMA and PEO-*b*-PBMA copolymers of various chain-length at different thermodynamic state points in the two chemical environments. From these observations, we hypothesize that the model is valid at any polymer concentration in THF over the well-defined range of temperatures studied here. As a case study and in order to test the transferability of the force-field to a different solvent, we have investigated the adsorption of a PEO-*b*-PBMA monolayer at the air-water interface showing that our model correctly predicts the experimental trend in the surface tension-molecular area isotherm for intermediate values of surface coverage. Thus, properties not easily accessible by experimental techniques such as the monolayer thickness and conformation of the single polymer chains within the monolayer can be obtained. Our reduced order CG models are computationally efficient as they can be used with the standard MARTINI time step of 20 fs, allowing one to overcome spatiotemporal limitations encountered in all-atom models and investigate interesting slow phenomena such as peptide and protein diffusion in model membranes and the phase and aggregation behaviour of the amphiphilic block-copolymers in selective solvents.

# Acknowledgement

The project leading to these results has received funding from the European Union's Horizon 2020 research and innovation programme under the Marie Skłodowska-Curie grant agreement No 676045 (MULTIMAT). The authors acknowledge the assistance given by IT Services and the use of the Computational Shared Facility at the University of Manchester.

# Supporting Information Available

The following files are available free of charge.

- SI-1: Probability distributions of bond lengths for the re-parameterized coarse-grained and atomistic systems of $PEO_6$-$b$-$PMMA_7$.

- SI-2: Probability distributions of angles for the re-parameterized coarse-grained and atomistic systems of $PEO_6$-$b$-$PMMA_7$.

- SI-3: Radial distribution functions of THF calculated using both atomistic and CG models.

# Graphical TOC Entry

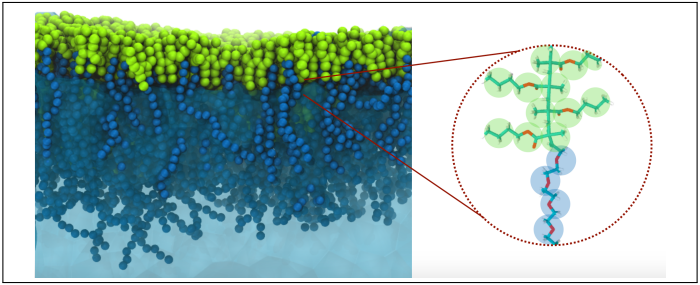

# SUPPORTING INFORMATION

# Transferable Coarse-Grained Model for Methacrylate-Based Copolymers

Gerardo Campos-Villalobos, Flor R. Siperstein, and Alessandro Patti*

*School of Chemical Engineering and Analytical Science, The University of Manchester, Sackville Street, M13 9PL, Manchester, UK*

E-mail: Alessandro.Patti@manchester.ac.uk

# 1. PEO$_6$-*b*-PMMA$_7$ Bond Length Distributions

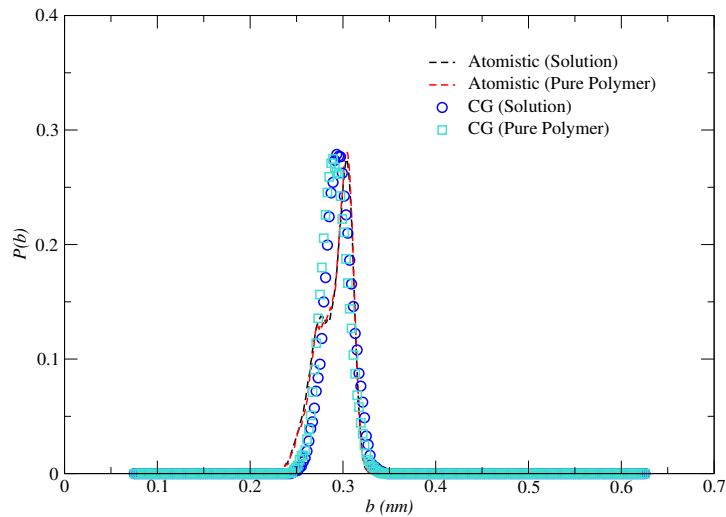

Figure 1: Probability distribution of bond length ($b$) between SP$_0$ and SP$_h$ segments sampled by atomistic models representing copolymer chains in THF (black dashed line) and pure polymer systems (red dashed line) along with the CG distributions (empty symbols).

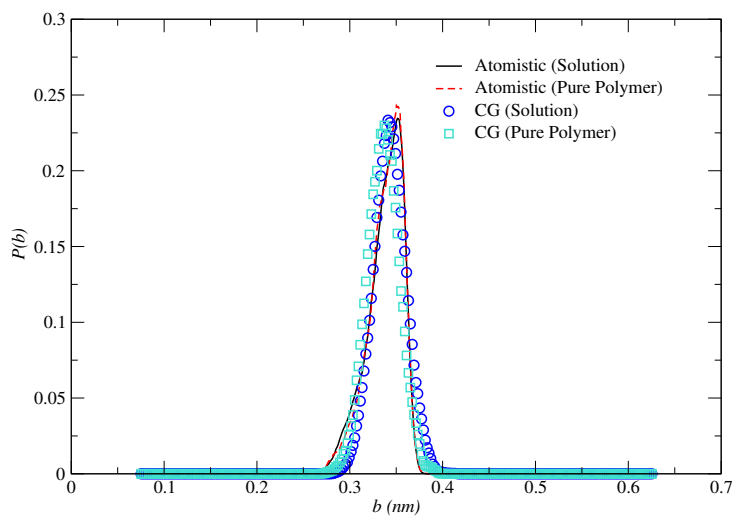

Figure 2: Probability distribution of bond length ($b$) between SP$_0$ and SP$_0$ segments sampled by atomistic models representing copolymer chains in THF (black dashed line) and pure polymer systems (red dashed line) along with the CG distributions (empty symbols).

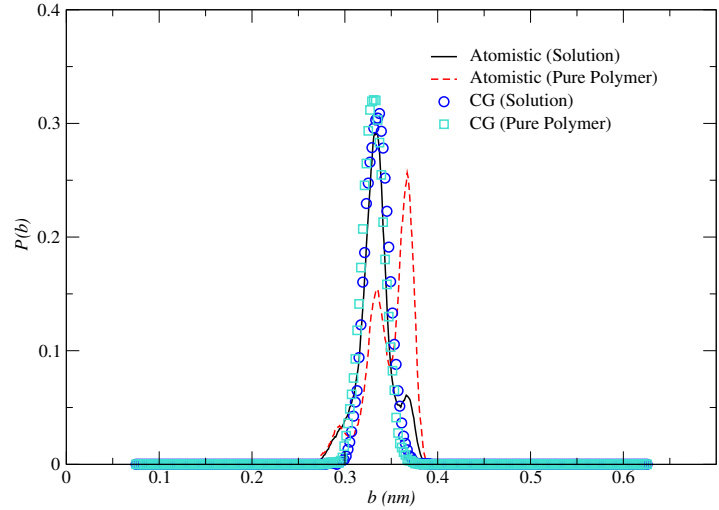

Figure 3: Probability distribution of bond length ($b$) between $SP_0$ and $SC_{1,B}$ segments sampled by atomistic models representing copolymer chains in THF (black dashed line) and pure polymer systems (red dashed line) along with the CG distributions (empty symbols).

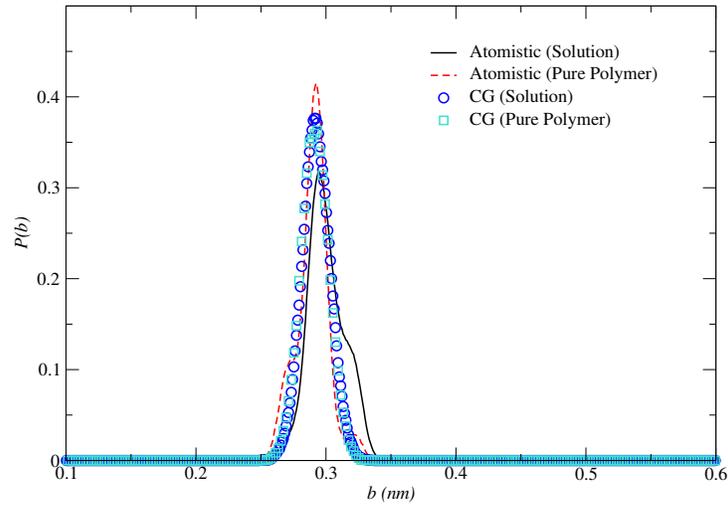

Figure 4: Probability distribution of bond length ($b$) between $SC_{1,B}$ and $SC_{1,B}$ segments sampled by atomistic models representing copolymer chains in THF (black dashed line) and pure polymer systems (red dashed line) along with the CG distributions (empty symbols).

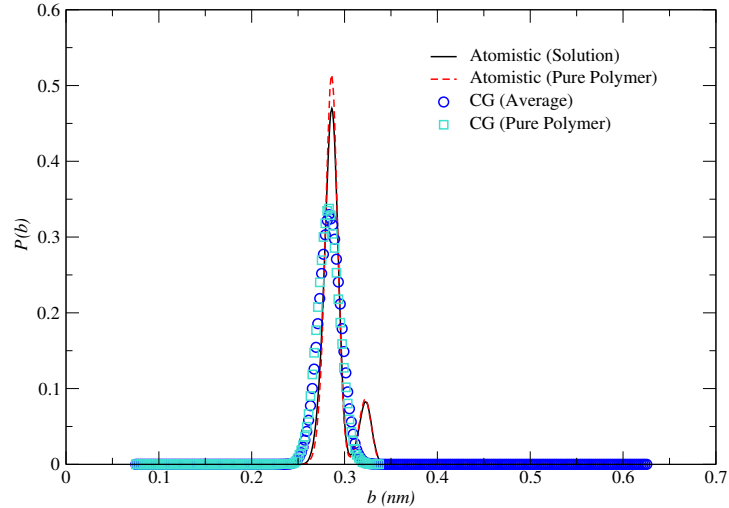

Figure 5: Probability distribution of bond length ($b$) between $SC_{1,B}$ and $N_a$ segments sampled by atomistic models representing copolymer chains in THF (black dashed line) and pure polymer systems (red dashed line) along with the CG distributions (empty symbols).

## 2. PEO$_6$-$b$-PMMA$_7$ Angle Distributions

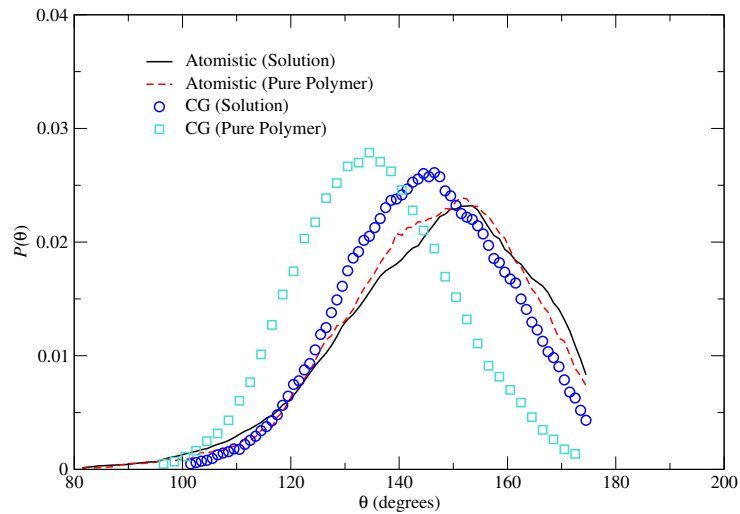

Figure 6: Probability distribution of angle ($\theta$) between $SP_h$-$SP_0$-$SP_0$ segments sampled by atomistic models representing copolymer chains in THF (black dashed line) and pure polymer systems (red dashed line) along with the CG distributions (empty symbols).

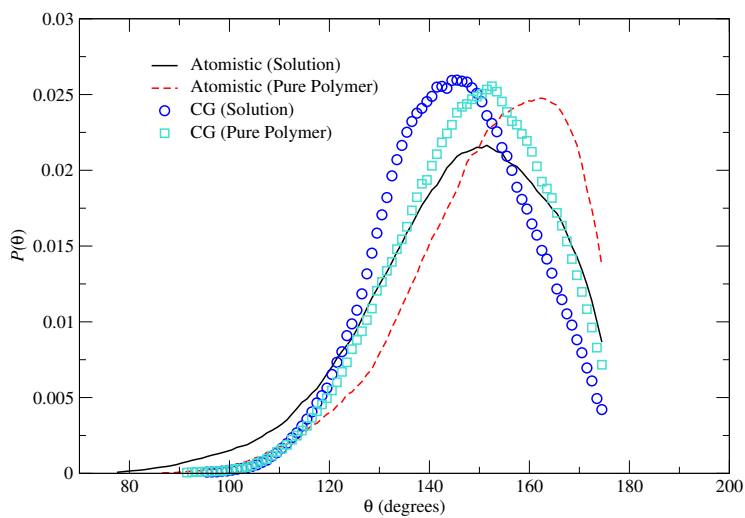

Figure 7: Probability distribution of angle ($\theta$) between $SP_0$-$SP_0$-$SP_0$ segments sampled by atomistic models representing copolymer chains in THF (black dashed line) and pure polymer systems (red dashed line) along with the CG distributions (empty symbols).

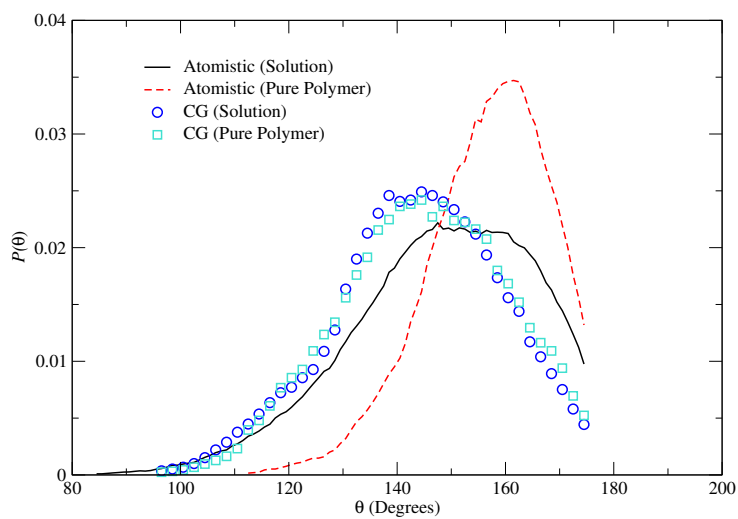

Figure 8: Probability distribution of angle ($\theta$) between $SP_0$-$SP_0$-$SC_{1,B}$ segments sampled by atomistic models representing copolymer chains in THF (black dashed line) and pure polymer systems (red dashed line) along with the CG distributions (empty symbols).

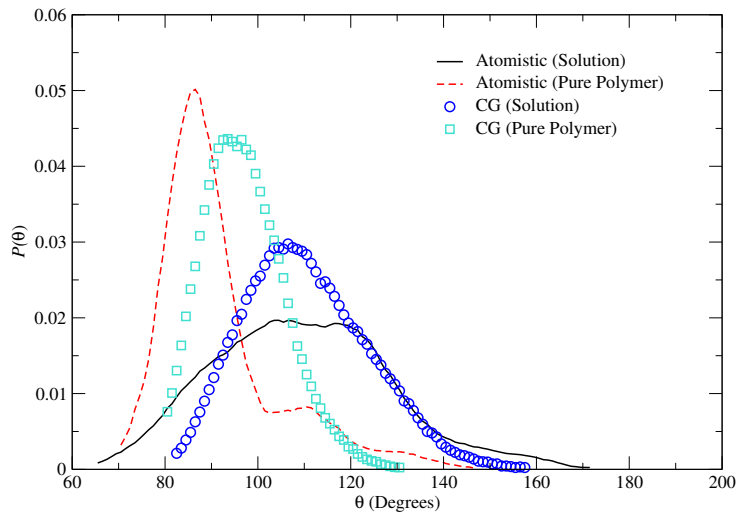

Figure 9: Probability distribution of angle ($\theta$) between $SP_0$-$SC_{1,B}$-$SC_{1,B}$ segments sampled by atomistic models representing copolymer chains in THF (black dashed line) and pure polymer systems (red dashed line) along with the CG distributions (empty symbols).

## 3. Equilibrium Structure of THF

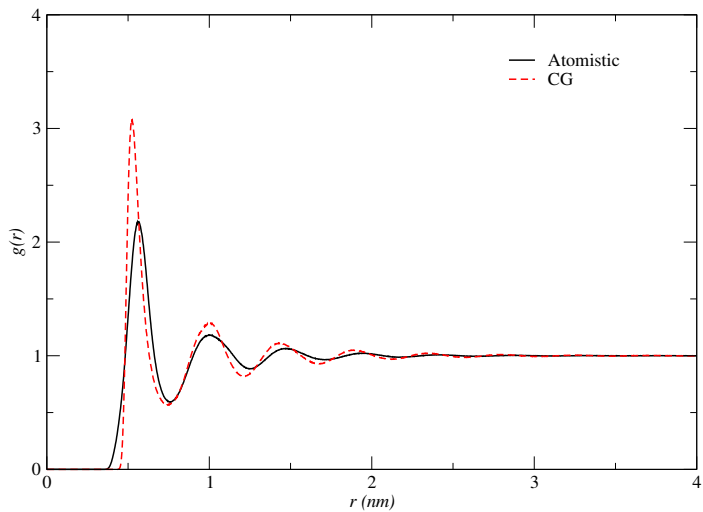

Figure 10: Radial distribution function (RDF) of THF obtained from atomistic (black solid line) and CG (red dashed line) simulations at 300 K and 1 bar. The atomistic RDF has been obtained from the trajectory of the center of mass of THF molecules.